\documentclass[reprint,preprintnumbers,nofootinbib,superscriptaddress,amsmath,amssymb,aps,prd]{revtex4-2}
\pdfoutput=1
\usepackage{graphicx,hyperref,braket,orcidlink, algorithm2e, amssymb, amsmath, dsfont,subfigure, pifont, slashed, xargs, multirow}
\graphicspath{{figs/}}
\hypersetup{
	bookmarks=true,
	unicode=true,
	pdftoolbar=true,
	pdfmenubar=true,
	pdffitwindow=false,
	pdfstartview={FitH},
	pdftitle={QOC},
	pdfauthor={J.Y.~Araz, M.~Spannowsky}, 
	pdfsubject={qVAE-anomaly},
	pdfcreator={J.Y.~Araz, M.~Spannowsky},
	pdfproducer={}, 
	pdfkeywords={},
	pdfnewwindow=true,
	colorlinks=true,
	linkcolor=blue,
	citecolor=olive,
	filecolor=magenta,
	urlcolor=cyan
}

\usepackage[colorinlistoftodos,prependcaption,textsize=tiny]{todonotes}
\newcommandx{\unsure}[2][1=]{\todo[linecolor=red,backgroundcolor=red!25,bordercolor=red,#1]{#2}}
\newcommandx{\change}[2][1=]{\todo[linecolor=blue,backgroundcolor=blue!25,bordercolor=blue,#1]{#2}}
\newcommandx{\info}[2][1=]{\todo[linecolor=blue,backgroundcolor=blue!25,bordercolor=blue,#1]{#2}}

\newcommand{\cmark}{\ding{51}}%
\newcommand{\xmark}{\ding{55}}%
\newcommand{\Ug}[1]{\hat{U}(#1)}%

\begin{document}

\title{The role of data embedding in quantum autoencoders for improved anomaly detection}

\author{Jack Y. Araz\orcidlink{0000-0001-8721-8042}}
\email{jackaraz@jlab.org}
\affiliation{Thomas Jefferson National Accelerator Facility, Newport News, VA 23606, USA}

\author{Michael Spannowsky\orcidlink{0000-0002-8362-0576}}
\email{michael.spannowsky@durham.ac.uk}
\affiliation{Institute for Particle Physics Phenomenology, Durham University, South Road, Durham DH1 3LE, United Kingdom}
\date{\today}

\preprint{JLAB-THY-24-4170, IPPP/24/60}

\begin{abstract}
The performance of Quantum Autoencoders (QAEs) in anomaly detection tasks is critically dependent on the choice of data embedding and ansatz design. This study explores the effects of three data embedding techniques—data re-uploading, parallel embedding, and alternate embedding—on the representability and effectiveness of QAEs in detecting anomalies. Our findings reveal that even with relatively simple variational circuits, enhanced data embedding strategies can substantially improve anomaly detection accuracy and the representability of underlying data across different datasets. Starting with toy examples featuring low-dimensional data, we visually demonstrate the effect of different embedding techniques on the representability of the model. We then extend our analysis to complex, higher-dimensional datasets, highlighting the significant impact of embedding methods on QAE performance.
\end{abstract}

\maketitle

\section{Introduction}

Anomaly detection, the process of identifying data points that deviate significantly from established patterns, is a critical task with applications spanning multiple domains. These include fraud detection in finance~\cite{ngai2011application}, fault detection in industrial systems~\cite{hodge2004survey}, and monitoring for cybersecurity threats~\cite{chandola2009anomaly}. Quantum computing, with its potential to provide exponential speedups over classical methods, offers novel and promising pathways to enhance machine learning algorithms~\cite{schuld2018supervised, biamonte2017quantum}, including those used for anomaly detection~\cite{liu2018quantum, Ngairangbam_2022, Duffy:2024zog, Araz:2022zxk}.

\begin{figure}
    \centering
    \includegraphics[width=\linewidth]{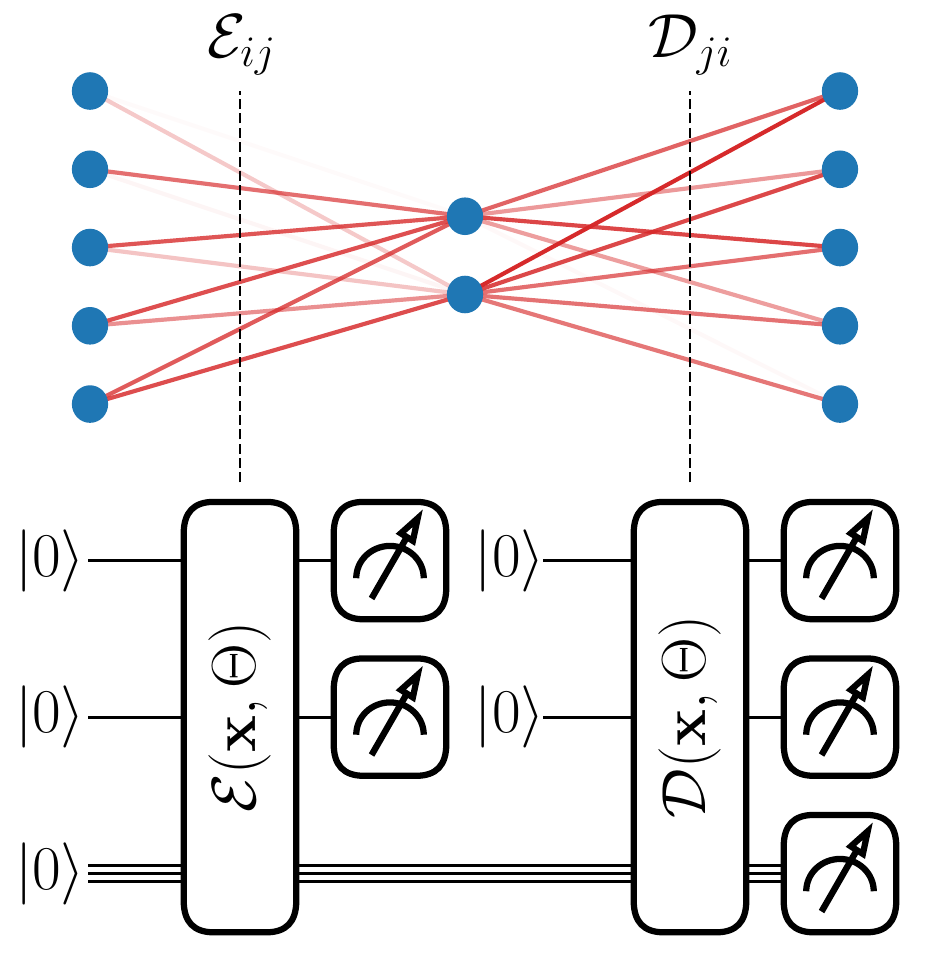}
    \caption{\it Schematic comparison between classical (top panel) and quantum (bottom panel) autoencoders. Encoder and decoder are labelled as $\mathcal{E}$ and $\mathcal{D}$, respectively.}
    \label{fig:clas-quant}
\end{figure}

Variational Autoencoders (VAEs)~\cite{kingma2022} have been extensively studied and applied in anomaly detection due to their ability to learn probabilistic representations of data. A VAE operates by encoding input data into a latent space and then reconstructing it; see the upper panel of Fig.~\ref{fig:clas-quant}. By learning the underlying distribution of normal data, VAEs can effectively flag anomalies as data points that deviate from this learned distribution. This approach has proven particularly useful in scenarios where the normal data distribution is complex and high-dimensional, making it difficult for simpler models to capture.

Building on the success of classical VAEs, Quantum Autoencoders (QAEs) have recently emerged as powerful tools for anomaly detection~\cite{Ngairangbam_2022, Sakhnenko:2021jme, 10.1007/978-981-99-8073-4_6, Tscharke:2023nzj, PhysRevD.107.016002, PhysRevA.99.052310, PhysRevB.108.165408, Duffy:2024zog}. QAEs utilize quantum circuits to learn compressed representations of normal data (see the lower panel of Fig.~\ref{fig:clas-quant}), potentially offering more efficient and accurate anomaly detection compared to their classical counterparts. However, the success of QAEs hinges on two crucial components: (1) the ansatz architecture, which determines the structure and expressive power of the quantum circuits, and (2) the data embedding method, which influences how well classical data is represented in the quantum framework.

Although initial studies have shown that QAEs can outperform classical autoencoders in specific anomaly detection tasks~\cite{Ngairangbam_2022}, there remains a significant gap in understanding how different ansatz designs and data embedding strategies impact the performance of QAEs. The choice of ansatz affects the model’s ability to generalize and capture complex data patterns~\cite{Bravo-Prieto:2020szj, Bilkis:2023aa}, while the embedding strategy plays a critical role in determining how effectively classical data can be mapped into quantum circuits. Techniques like data reuploading and parallel embedding offer different trade-offs in terms of resource efficiency and representational power~\cite{P_rez_Salinas_2020, Schuld:2020enb}.

Previous research has primarily focused on comparing various variational circuit design strategies to improve QAE performance or utilizing hybrid techniques to maximize the limited resources available~\cite{Duffy:2024zog, Sakhnenko:2021jme}. Other studies have explored semi-supervised techniques~\cite{Tscharke:2023nzj} or employed hybrid quantum-classical schemes, such as embedding affine transformations of the data with trainable parameters~\cite{Bravo-Prieto:2020szj}. However, to our knowledge, there has been no comprehensive analysis of the impact of different quantum embedding methods—specifically data re-uploading and parallel embedding, as introduced in Refs.~\cite{P_rez_Salinas_2020, Schuld:2020enb}—within the context of anomaly detection using QAEs.

This paper aims to fill this gap by providing a detailed comparative analysis of various data embedding methods in the context of QAE-based anomaly detection. We systematically explore how different embedding structures influence the capacity and performance of QAEs and evaluate their effectiveness in optimizing the representation of classical data within a quantum space. Our study highlights the strengths and limitations of each embedding approach and offers practical insights into the best practices for designing robust and efficient QAEs for anomaly detection applications. 

This study is organized as follows: Section~\ref{sec:qve} provides a brief introduction to quantum autoencoders, setting the foundation for understanding their role in anomaly detection. In Section~\ref{sec:embed}, we delve into the various data embedding techniques and ansatz architectures employed in our experiments. The results are presented and analyzed in Section~\ref{sec:results}, with a focus on 2D datasets in Section~\ref{sec:2d-dat} and high-dimensional datasets in Section~\ref{sec:high-dim-dat}. Finally, our conclusions are discussed in Section~\ref{sec:conclusion}, summarizing the key findings of the study.

\section{Quantum Autoencoder}\label{sec:qve}
\begin{figure}[t]
    \centering
    \includegraphics[width=\linewidth]{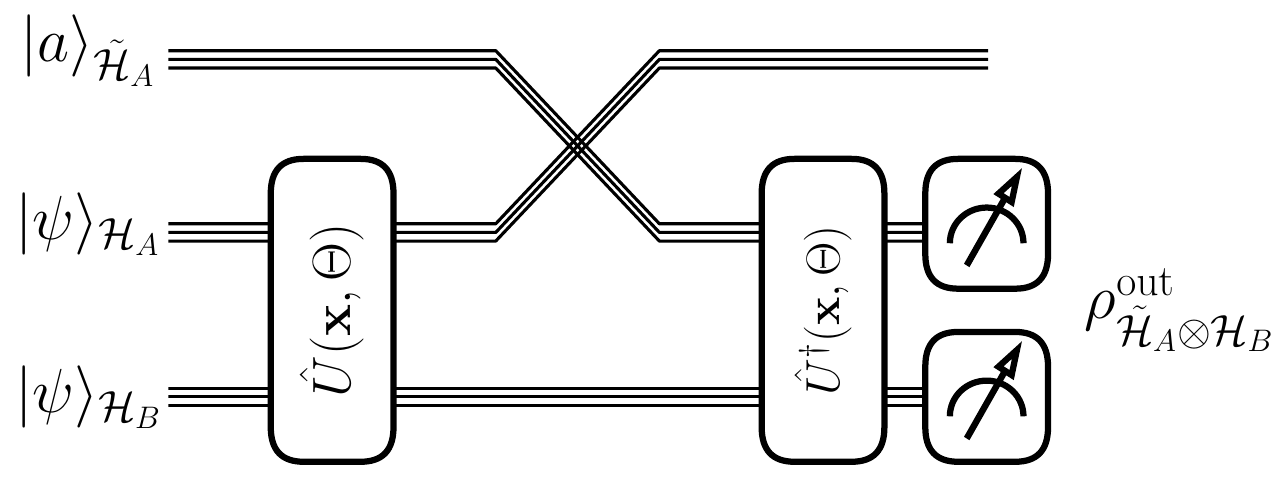}
    \caption{\it Schematic representation of a quantum autoencoder circuit.}
    \label{fig:qae-psi}
\end{figure}

A classical VAE consists of two main parts: an encoder ($\mathcal{E}$) and a decoder ($\mathcal{D}$). The encoder embeds the input data into an ansatz and compresses it into a lower-dimensional latent space. The decoder then takes this compressed latent space as input and expands it back to the original dimensionality of the input data. By optimizing the distance or difference between the input data and the output from the network, the model can effectively capture the characteristics of the input. If the model is trained on a specific dataset, it can be used as a measure of anomaly to differentiate this dataset from others. The upper panel of Fig.~\ref{fig:clas-quant} presents a schematic representation of the VAE, where a five-dimensional input is compressed into a two-dimensional latent space by the encoder and then reconstructed by the decoder. The encoder ($\mathcal{E}_{ij}$) and decoder ($\mathcal{D}_{ji}$) are represented as trainable network ansätze.

Similar to its classical counterpart, the QAE comprises an encoder ($\mathcal{E}$) and a decoder ($\mathcal{D}$) separated by an information bottleneck. In this case, however, the encoder and decoder are implemented as sets of unitary transformations, i.e., quantum gates, that transform the input into a quantum state. Unlike classical autoencoders, where parts of the feature space can be discarded, unitary transformations preserve the entire Hilbert space. To address this, certain states are traced out to achieve the desired dimensionality reduction. The schematic representation of this quantum circuit is shown in the bottom panel of Fig.~\ref{fig:clas-quant}, where $\mathbf{x}$ denotes the inputs, and $\Theta$ represents the trainable parameters of the unitary transformations. A mid-circuit measurement has represented the dimensionality reduction.

\begin{figure}[!t]
    \centering
    \includegraphics[width=0.9\linewidth]{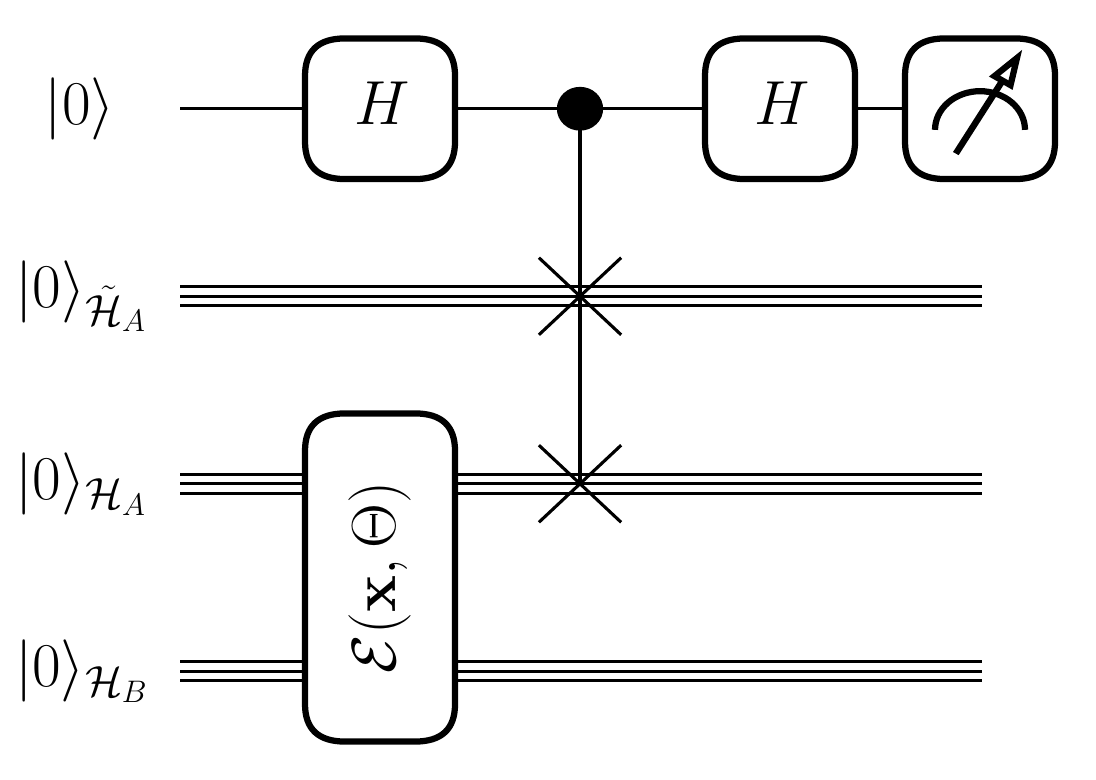}
    \caption{\it Schematic representation of the quantum autoencoder circuit. $\mathcal{E}(\mathbf{x}, \Theta)$ represents the encoder circuit, $H$ is the Hadamard gate, and $\mathcal{H}_i$ represents different Hilbert spaces.}
    \label{fig:qvae}
\end{figure}

A more efficient quantum circuit can be implemented to optimise the training process. The transformation into a latent space and subsequent reconstruction can be achieved by inverting the encoder unitary and replacing reference states with a set of trash states $|a\rangle_{\tilde{\mathcal{H}}_A}\equiv|a\rangle$, typically initialised to $|0\rangle^{\otimes {\rm dim}[\tilde{\mathcal{H}}_A]}\equiv |0\rangle_{\tilde{\mathcal{H}}_A}$. This circuit is depicted in Fig.~\ref{fig:qae-psi}. To optimise the unitary transformation, the goal is to minimise the difference between the input state $|\psi\rangle \equiv |\psi\rangle_{\mathcal{H}_A}\otimes |\psi\rangle_{\mathcal{H}_B}$ and the output state $\rho^{\rm out}\equiv\rho^{\rm out}_{\tilde{\mathcal{H}}_A \otimes \mathcal{H}_B}$, which can be done by measuring the fidelity between these two states, $\lVert \langle \psi | \rho^{\rm out} | \psi\rangle\rVert^2$. Notably, since the decoder is the inverse of the encoder, the circuit can be further simplified using a SWAP test~\cite{PhysRevLett.87.167902}, which efficiently computes the projection between two states—in this case, between $|a\rangle$ and $|\psi\rangle_{\mathcal{H}_A}$. Perfect reconstruction is achieved when $|a\rangle = |\psi\rangle_{\mathcal{H}_A}$, i.e. when the fidelity is 1. This simplified circuit is illustrated in Fig.~\ref{fig:qvae}, where $H$ represents the Hadamard gate, and the states are swapped via a controlled SWAP gate. The probability of successful reconstruction can be measured by projecting the encoder state in the reduced Hilbert space on trash states,
\begin{eqnarray}
    p(\mathbf{x}, \Theta) = \left\lVert {\rm Proj}\left[ |0\rangle_{\tilde{\mathcal{H}}_A}, |\psi\rangle_{\mathcal{H}_A}\right] \right\rVert^2\ ,
\end{eqnarray}
where $|\psi\rangle \equiv \mathcal{E}(\mathbf{x}, \Theta) (| 0\rangle_{\mathcal{H}_A} \otimes |0\rangle_{\mathcal{H}_B})$. Both data embedding and the parameterised unitary transformations are encapsulated in $\mathcal{E}(\mathbf{x}, \Theta)$, which will be detailed in the following section. 
Using the fidelity as a probability distribution of the state being reconstructed properly, we can construct mean negative log-probability as our cost function to minimise,
\begin{eqnarray}
\mathcal{C}(\mathbf{X},\Theta) = -\frac{1}{N}\sum_i^N\log p(\mathbf{x}_i,\Theta)\ ,
\end{eqnarray}
to learn the likelihood distribution of the features. Here, $N$ represents the number of samples and $\mathbf{X}$ includes $N$ number of samples. Previous studies have also used cost functions like 1-fidelity; however, we did not observe any significant difference in the results. We will use the negative log-probability as an anomaly measure to test our algorithm to differentiate between normal and anomalous samples. In the following section, we will discuss the data embedding procedure and the choice of ansatz for $\mathcal{E}(\mathbf{x}, \Theta)$.

\section{Data embedding and the choice of ansatz}\label{sec:embed}

Many different data embedding strategies have been developed for quantum machine learning applications. Two main options are amplitude embedding and angle embedding. In this study, we focus on the latter, but it should be noted that the former is more qubit efficient, but it has been suggested that amplitude embedding may be more prone to barren plateaus~\cite{Huang:2021aa, PRXQuantum.2.040337}, which is out of scope of this study.

Angle embedding rotates each qubit by an angle defined by the input data, effectively mapping real data on a quantum state, \( R_P(x_i)|0\rangle \), where \( P \) is a Pauli operator \( (P \in \{X, Y, Z\}) \). The \( R_P \) operation for a set of features, $x_i\in\mathbf{x}$, is expressed as:
\begin{eqnarray}
    S_{\rm st}(\mathbf{x}) =  \bigotimes^n_{i=0} e^{-ix_i P_{(i)}} \ ,  \label{eq:angle-emb}
\end{eqnarray}
where \( \otimes \) denotes the Kronecker product and the subscript \( (i) \) indicates the qubit on which the operation is applied. In this study, we will explore the effects of expanding this embedding. Our previous work~\cite{Araz:2021zwu} demonstrated that embedding data within a multidimensional hypersphere can enhance the representability of the ansatz, a strategy we will also explore here. We will test this with two embedding methods. 

First, we introduce parallel embedding,
\begin{eqnarray}
    S_{\rm P}(\mathbf{x}) =  \bigotimes^n_{i=0} \left(e^{-ix_i Y_{(2i)}} \otimes e^{-ix_i Y_{(2i+1)}} \right)\ , \label{eq:p}
\end{eqnarray}
where \( P \) is fixed to \( Y \) (though any Pauli operator can be used). Here, a single feature is embedded across two adjacent qubits, representing the number of features with $n$. Due to computational constraints, we limit this implementation to two qubits, but it can be extended to more. 

The second method is parallel and alternate embedding,
\begin{eqnarray}
    S_{\rm A}(\mathbf{x}) =  \bigotimes^n_{i=0}\left( e^{-i x_i Y_{(2i)}} \otimes e^{-ix_i X_{(2i+1)}}\right)\ , \label{eq:pa}
\end{eqnarray}
where the second Pauli operator is switched to \( X \). These methods aim to enhance data representation on a higher-dimensional manifold. A generalised version of these can be formulated as follows;
\begin{eqnarray}
    S_{\rm G}(\mathbf{x}) = \bigotimes_{i=0}^n \left[ \bigotimes_{j=0}^d e^{-i x_i P_{(di+j)}} \right]
\end{eqnarray}
where $d$ is the dimensionality of the embedding and Pauli operator $P$ can be cycled between $X$, $Y$ and $Z$. This indicates that the number of qubits required for this embedding is $d\times n$ while the depth of the quantum circuit remains the same as Eq.~\eqref{eq:angle-emb}.

We define the parameterised unitary ansatz to form the encoder following data embedding. In this study, we use strongly entangling layers~\cite{Schuld:2018ahn}, with each layer denoted as \( \Ug{\Theta_i} \), where \( i \) is the layer index and \( \Theta_i \) represents the parameters of that layer. Unless otherwise specified, these layers consist of parameterised rotation gates, typically one \( R_Y(\theta \in \Theta_i) \) per qubit. The rotation gates are followed by CNOT gates, which entangle each qubit with every other qubit.


Finally, we construct the encoder by combining the parameterised unitary with data embedding. We will use two approaches. The first, a standard method employed in many studies, applies data embedding followed by parameterised unitaries:
\begin{eqnarray}
    \mathcal{E}_{\rm st}(\mathbf{x}, \Theta) = S(\mathbf{x}) \prod^L_{i=0} \hat{U}(\Theta_i)\ .\label{eq:encoder-st}
\end{eqnarray}
Here, \( L \) represents the number of layers.

The second approach uses data reuploading~\cite{P_rez_Salinas_2020, Schuld:2020enb}, which encodes data repeatedly in every layer, enhancing representation. In this method, data is repeatedly embedded in each layer:
\begin{eqnarray}
    \mathcal{E}_{\rm R}(\mathbf{x}, \Theta) = \hat{U}(\Theta_0) \prod^L_{i=1} S(\mathbf{x}) \hat{U}(\Theta_i) \ . \label{eq:encoder-reupload}
\end{eqnarray}
Notice that each $\hat{U}(\Theta_i)$ represents a single layer of strongly entangled ansatz. The data re-uploading method has been shown to represent Fourier expansion for input data, which can be used to approximate any function, achieving a universal quantum encoding~\cite{Schuld:2020enb}. It is important to note that Eq.~\eqref{eq:encoder-reupload} increases the depth of a given quantum circuit by one per layer compared to Eq.~\eqref{eq:encoder-st}, which makes it more prone to errors due to the limited coherence time.

\section{Results}\label{sec:results}

\begin{table}[!t]
    \centering
    \setlength\tabcolsep{5pt}
    \begin{tabular}{|l||c|c|c|}
    \hline
        Dataset & \# Training & \# Validation & \# Test \\\hline
        Moons & 40000 & 10000 & 5000\\
        S Curve & 8000 & 2000 & 5000 \\
        Circle & 31000 & 8000 & 5000\\
        Donut &  40000 & 10000 & 5000\\\hline
        Creditcard Fraud & 225177 & 56863 & 2767 \\\hline
    \end{tabular}
    \caption{\it Number of training, validation and test sample sizes for each dataset.}
    \label{tab:dataset-size}
\end{table}

In this section, we present our results for two groups of datasets. In section~\ref{sec:2d-dat}, we discuss the results for 2D datasets, which allows for the visualisation of the methods used. In section~\ref{sec:high-dim-dat}, we apply our approach to a higher-dimensional dataset to demonstrate that our findings hold in higher dimensions.

We fixed a certain set of hyperparameters for all the results presented below. Each model was trained for 500 epochs with a batch size of 100. We used \textsc{Adam} optimiser~\cite{adam} with an initial learning rate of 0.1. We applied an exponential decay scheduler, which reduced the learning rate every 100 epochs with a decay rate of 0.5. Trainable parameters were initialized using a uniform distribution between \([- \pi, \pi]\). If overtraining occurred, we stopped the training and used the parameters from the point before overtraining began. The number of samples in the training, validation, and test sets for each dataset is provided in Table~\ref{tab:dataset-size}. For the simulations, we used \textsc{PennyLane} (version 0.35.1)~\cite{bergholm2020pennylane} along with \textsc{Jax} (version 0.4.30)~\cite{jax2018github} and CUDA (version 12.5)~\cite{cuda}.
\begin{figure*}[!t]
    \centering
    \includegraphics[width=\linewidth]{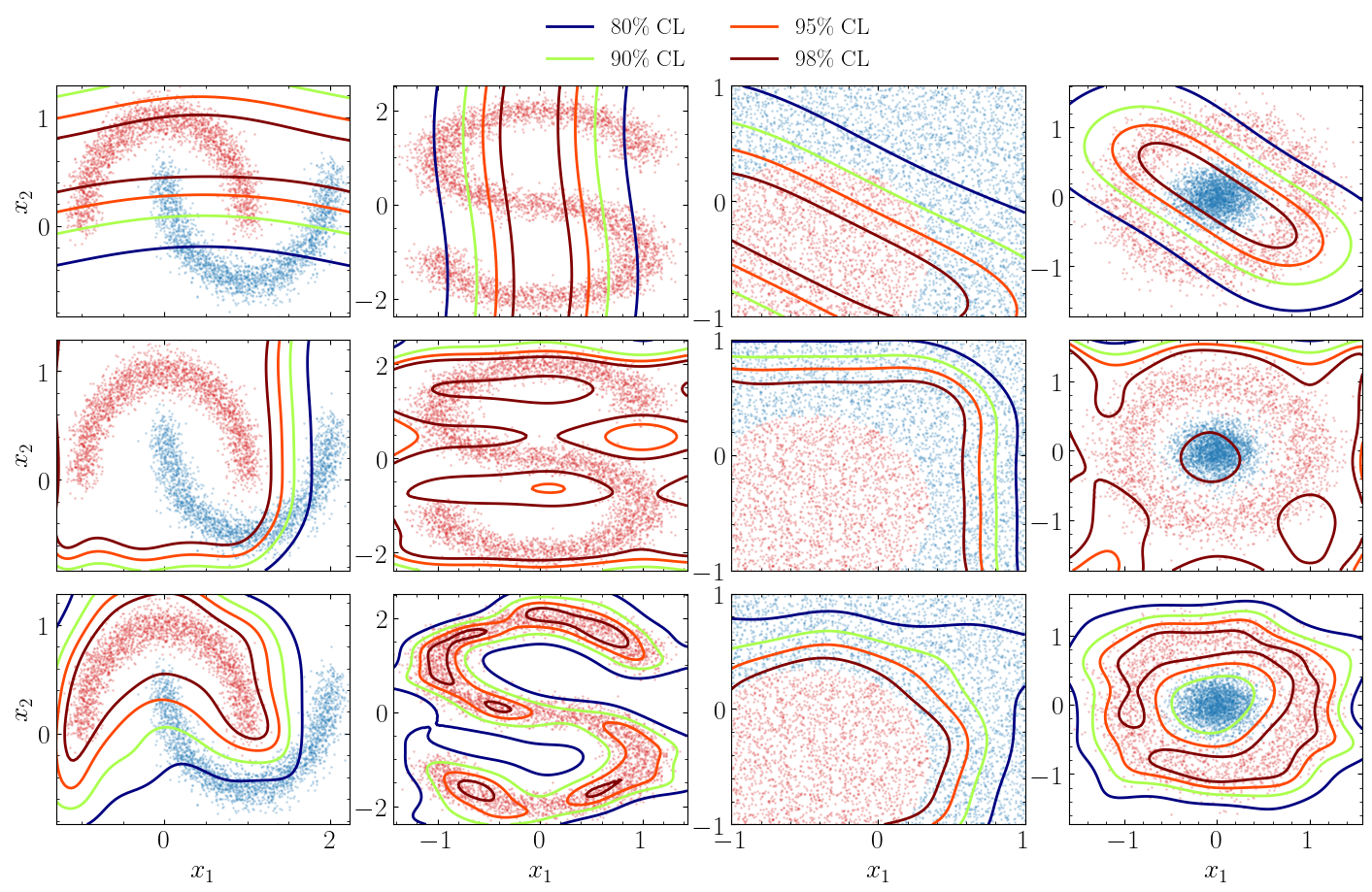}
    \caption{\it Selected results from 2D datasets. From left to right, the image shows moons, s-curve, circle and doughnut datasets and from top to bottom, it shows the models $M_1$, $M_5$ and $M_9$.}
    \label{fig:results-2D}
\end{figure*}

\subsection{Low dimensional datasets}\label{sec:2d-dat}

To visually examine the effects of data embedding in anomaly detection, we first utilised standard 2D datasets from \textsc{Scipy} (version 1.10.1)~\cite{2020SciPy-NMeth}. The sample sizes for each dataset are presented in the upper section of Table~\ref{tab:dataset-size}.

\begin{table}[!h]
    \centering
    \begin{tabular}{|c||c|c|c|c|c|c|c|c|c|}
    \cline{7-10}
    \multicolumn{6}{c|}{} & \multicolumn{4}{c|}{AUC} \\\hline
        Models & R & P & A & L & Comp. & Moons & S Curve & Circle & Doughnut\\\hline\hline
         $M_1$ & \xmark & \xmark & \xmark & 4 & Y      & 0.916 & 0.601 & 0.919 & 0.047 \\
         $M_2$ & \cmark & \xmark & \xmark & 4 & Y      & 0.904 & 0.861 & 0.999 & 0.988 \\
         $M_3$ & \cmark & \cmark & \xmark & 4 & Y      & 0.974 & 0.743 & 0.982 & 0.797 \\
         $M_4$ & \cmark & \xmark & \cmark & 4 & Y      & 0.950 & 0.889 & 0.991 & 0.843 \\
         $M_5$ & \cmark & \xmark & \xmark & 8 & Y      & 0.845 & 0.776 & 0.915 & 0.914 \\
         $M_6$ & \cmark & \xmark & \cmark & 8 & Y      & 0.996 & 0.924 & 0.984 & 0.895 \\
         $M_7$ & \cmark & \xmark & \cmark & 4 & YXY    & 0.995 & 0.836 & 0.879 & 0.936 \\
         $M_8$ & \cmark & \xmark & \cmark & 6 & YXY    & 1     & 0.945 & 0.995 & 0.937 \\
         $M_9$ & \cmark & \xmark & \cmark & 8 & YXY    & 0.999 & 0.984 & 0.998 & 0.974 \\\hline
    \end{tabular}
    \caption{\it Selected models and corresponding AUC value for each 2D dataset. From left to right, the columns represent data-reuploading (R), parallel encoding (P), alternate encoding (A), number of layers (L) and the composition of the ansatz (comp.). The rest of the columns show the datasets.}
    \label{tab:model-2d}
\end{table}

We employed nine benchmark models to investigate the impact of different data embedding techniques. The properties of these models, along with the area under the Receiver Operating Characteristic (ROC) curve (denoted as AUC) for each dataset, are detailed in Table~\ref{tab:model-2d}. The table is organised as follows: the first column (labelled R) indicates whether the model uses the standard encoder (Eq.~\eqref{eq:encoder-st}) or the data-reuploading encoder (Eq.~\eqref{eq:encoder-reupload}). The subsequent columns specify whether parallel (P) or alternate (A) embedding was applied. The number of layers used (4, 6, or 8) is listed in the column labelled L. The composition (comp.) column details the set of rotation gates used in the strongly entangling layers, $\hat{U}(\Theta)$, where $Y$ represents $R_Y(\theta)$, and $YXY$ denotes a composition of $R_Y(\theta_1)R_X(\theta_2)R_Y(\theta_3)$ for each qubit. In addition to AUC, we used accuracy values computed at 60\% and 80\% true positive rate (TPR) (or signal efficiency), presented in Table~\ref{tab:2d-acc} for each model and dataset.

\begin{table}[!h]
    \centering
    \setlength\tabcolsep{6pt}
    \begin{tabular}{|c|c|c|c|c|c|}
    \cline{3-6}
        \multicolumn{2}{c|}{} & \multicolumn{4}{c|}{Accuracy}\\\hline
        Model & TPR & Moons & S Curve & Circle & Doughnut  \\\hline\hline
         \multirow{2}{*}{$M_1$}& 60\% & 79\% & 50\% & 76\% & 30\% \\
                               & 80\% & 83\% & 43\% & 83\% & 40\% \\\hline
         \multirow{2}{*}{$M_2$}& 60\% & 78\% & 76\% & 76\% & 80\% \\
                               & 80\% & 84\% & 77\% & 88\% & 89\% \\\hline
         \multirow{2}{*}{$M_3$}& 60\% & 80\% & 72\% & 76\% & 71\% \\
                               & 80\% & 88\% & 58\% & 88\% & 76\% \\\hline
         \multirow{2}{*}{$M_4$}& 60\% & 80\% & 79\% & 76\% & 72\% \\
                               & 80\% & 89\% & 80\% & 88\% & 79\% \\\hline
         \multirow{2}{*}{$M_5$}& 60\% & 79\% & 73\% & 76\% & 77\% \\
                               & 80\% & 76\% & 62\% & 88\% & 84\% \\\hline
         \multirow{2}{*}{$M_6$}& 60\% & 80\% & 81\% & 76\% & 75\% \\
                               & 80\% & 90\% & 85\% & 86\% & 84\% \\\hline
         \multirow{2}{*}{$M_7$}& 60\% & 80\% & 81\% & 76\% & 77\% \\
                               & 80\% & 90\% & 74\% & 88\% & 86\% \\\hline
         \multirow{2}{*}{$M_8$}& 60\% & 80\% & 82\% & 76\% & 77\% \\
                               & 80\% & 90\% & 86\% & 88\% & 86\% \\\hline
         \multirow{2}{*}{$M_9$}& 60\% & 80\% & 83\% & 76\% & 79\% \\
                               & 80\% & 90\% & 90\% & 88\% & 88\% \\\hline
    \end{tabular}
    \caption{\it Accuracy values for each model presented in Table~\ref{tab:model-2d} for two true positive rate (TPR or signal efficiency) points at 60\% and 80\%.}
    \label{tab:2d-acc}
\end{table}

The selected results are presented in Fig.~\ref{fig:results-2D}. The results for moons, s-curve, circle, and doughnut datasets are shown from left to right. Only models 1,5 and 9 are selected to be shown in Fig.~\ref{fig:results-2D}, which are positioned from top to bottom. Decision boundaries are presented using confidence intervals at 80\%, 90\%, 95\%, and 98\%, which are shown with blue, green, orange and dark red, respectively. These intervals were computed using a \(\chi^2\) distribution,
\[
    \chi^2(x) = -2\log\frac{p(x, \Theta)}{p(\hat{x}, \Theta)}\ ,
\]
where \(\hat{x}\) represents the grid points that maximize the probability within the region shown in each plot for a fixed set of parameters $\Theta$. The \(\chi^2\) values were calculated using grid data generated within the region presented in each plot, independent of the actual data. Each model was trained on the data represented by red dots, and AUC (shown in Table~\ref{tab:model-2d}) and accuracy (shown in Table~\ref{tab:2d-acc}) values were computed based on the data shown with blue dots. Training, validation, and test data were generated independently using different random seeds.\footnote{AUC and accuracy values for the s-curve dataset are calculated with respect to pseudo-data generated outside of the s-curve region, where the radial distance to any red point has been chosen to be greater than 0.07 au.}

We observe that the AUC values do not necessarily indicate the successful reconstruction of the provided dataset. For instance, model $M_1$ for the moons dataset resulted in a $0.916$ AUC value due to the separation between the red and blue datasets, where a simple line can easily separate them. However, this does not indicate a good representation of the data with which the model has been trained. Similar behaviour can be observed in the circle dataset as well. Throughout each dataset, we observe significant improvement in the AUC value and accuracy of the model once different data embedding methods have been implemented. For instance, circle data can be perfectly represented using the data reuploading method with only four layers. Notice that it may even be possible to achieve such results with fewer layers, but the goal of this study is not to find the most efficient method but to compare different data embedding methods.

It is important to emphasize that due to the simplicity of the 2D datasets, implementing various embedding strategies without sufficiently large variational ansatz has been observed to reduce the model's performance for specific instances. This is because the ansatz is not expressive enough to suppress the over-expressiveness of the embedding. For instance, we observe that four-layer data reuploading is highly efficient in representing circular data; any addition increases the complexity of the training.

Overall, all the datasets used for this exercise resulted in improved representation when data reuploading, parallel, or alternate embedding was used. However, since 2D datasets are highly simple, we investigate a higher-dimensional dataset in the following section.

\begin{table}
    \centering
    \setlength\tabcolsep{6pt}
    \begin{tabular}{|c|c|c|c|c|c|c|}
    \cline{3-7}
        \multicolumn{2}{c|}{} & \multicolumn{5}{c|}{Accuracy}\\\hline
        L/\#Ref & TPR & Base & Base A & R & RP & RA \\\hline\hline
         \multirow{2}{*}{4/1}& 60\% & 81\% & 88\% & 91\% & 91\% & 91\%\\
                             & 80\% & 55\% & 85\% & 85\% & 84\% & 85\%\\\hline
         \multirow{2}{*}{4/3}& 60\% & 94\% & 90\% & 90\% & 90\% & 91\% \\
                             & 80\% & 73\% & 83\% & 88\% & 89\% & 89\%\\\hline
         \multirow{2}{*}{6/1}& 60\% & 82\% & 90\% & 91\% & 91\% & 91\%\\
                             & 80\% & 54\% & 86\% & 90\% & 87\% & 87\%\\\hline
         \multirow{2}{*}{6/3}& 60\% & 91\% & 90\% & 91\% & 91\% & 91\%\\
                             & 80\% & 72\% & 87\% & 87\% & 88\% & 88\%\\\hline
         \multirow{2}{*}{8/1}& 60\% & 91\% & 91\% & 92\% & 91\% & 91\%\\
                             & 80\% & 55\% & 85\% & 90\% & 87\% & 87\%\\\hline
         \multirow{2}{*}{8/3}& 60\% & 91\% & 90\% & 91\% & 90\% & 90\%\\
                             & 80\% & 71\% & 87\% & 85\% & 89\% & 89\%\\\hline
    \end{tabular}
    \caption{\it Accuracy values for each model presented for credit card fraud data. The first column shows a number of layers (L) and a number of reference states (\#Ref) used in the model, followed by true positive rate (TPR) benchmarks at 60\% and 80\%.}
    \label{tab:multidim-acc}
\end{table}

\subsection{High dimensional datasets}\label{sec:high-dim-dat}
\begin{figure*}[!t]
    \centering
    \includegraphics[width=0.95\linewidth]{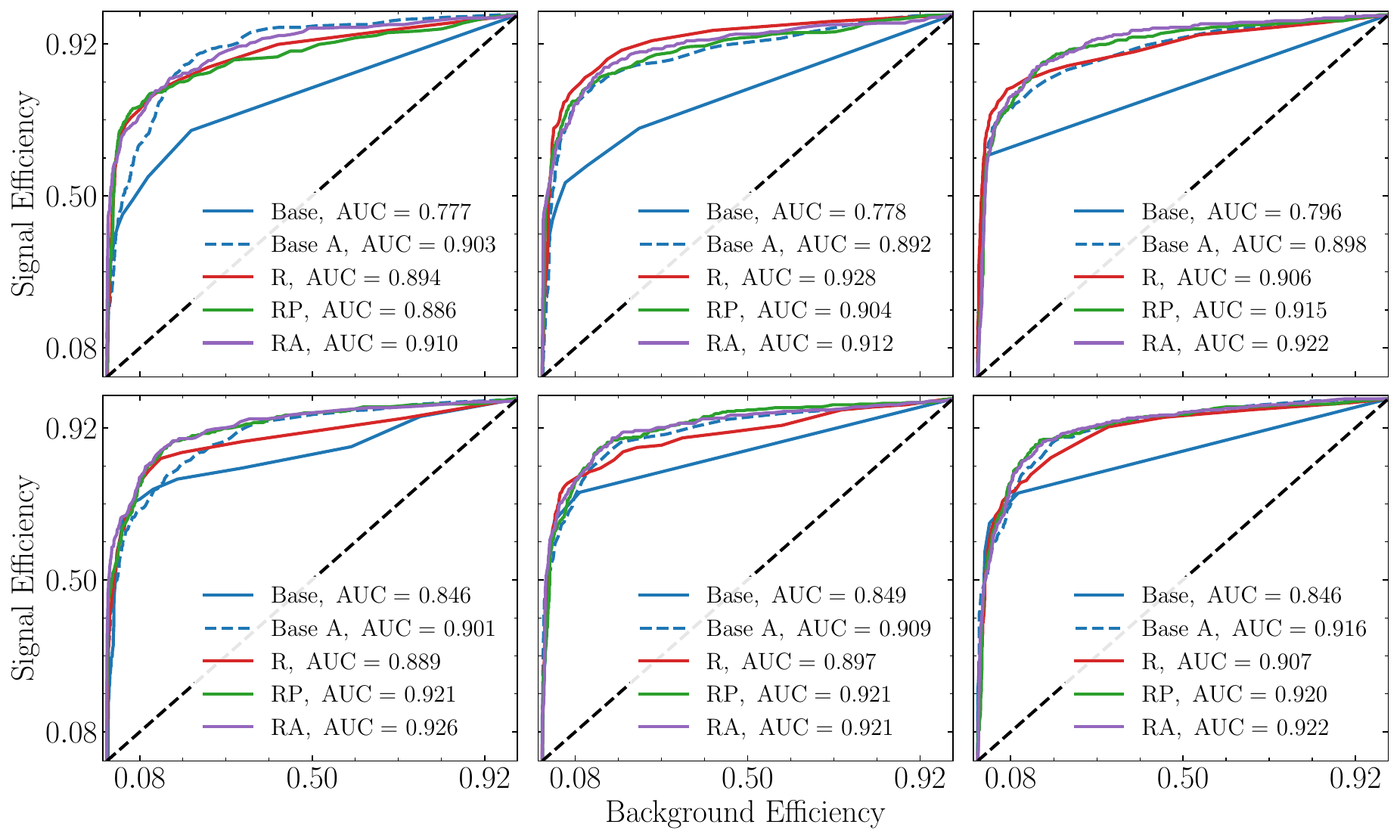}
    \caption{\it The ROC curves show the results for credit card fraud data, where rows represent 1 and 3 reference states used in the model and from left to right, each plot shows 4, 6, and 8-layer models. Letters are used in the legend to represent data-reuploading (R), parallel embedding (P) and alternate embedding (A). The dashed black line represents the random choice.}
    \label{fig:result-creditcard}
\end{figure*}

To evaluate the effectiveness of various data embedding approaches in a general context, we used Kaggle's credit card fraud dataset. The data was standardized using Scipy's MinMaxScaler, scaling it to the range $[-\pi, \pi]$. The 492 fraudulent cases were isolated from the rest of the dataset and reserved exclusively for testing. We focused on the first five columns of the dataset, labelled as $V_i$ in the original data. A link to the dataset is provided in Section~\ref{sec:reproduce}.

Our results are organized based on several benchmarks, including the number of layers—4, 6, and 8—and the number of reference states—1 and 3. Additionally, we included a reference benchmark, labelled 'base,' representing the standard embedding approach using Eq.~\eqref{eq:encoder-st} with angle embedding (Eq.~\eqref{eq:angle-emb}). The other models are identified by letters: R for data reuploading, P for parallel embedding, and A for alternate embedding. Reference states were selected as the last one or three qubits where the data was embedded, regardless of the embedding type. For example, two of the last three qubits carry the data in parallel embedding.

Figure~\ref{fig:result-creditcard} presents the ROC curves for each benchmark. The first row shows benchmarks with one reference state, while the second row shows benchmarks with three reference states. The plots are arranged from left to right according to the number of layers—4, 6, and 8, respectively. We observe that the base model is significantly affected by both the number of layers and reference states. However, models utilizing different embedding methods demonstrate considerable robustness to these variations. When alternative data embedding or re-uploading techniques were applied, no significant improvements were observed with changes in the number of layers or reference states. The results fluctuated within a few per cent for each benchmark.

Additionally, we present the accuracy for each benchmark in Table~\ref{tab:multidim-acc}, focusing on two True Positive Rate (TPR) benchmarks at 60\% and 80\%. A significant drop in accuracy between the 60\% and 80\% benchmarks was observed in all the base models, whereas this drop was significantly reduced in all other models. This indicates that these embedding strategies enable a more robust ansatz for anomaly detection.

The key takeaway from this exercise is the substantial improvement achieved through enhanced data embedding. Across all benchmarks, significant gains were observed when data reuploading or parallel and alternate embedding were used. This suggests that enhanced data embedding is crucial for improving the representability of the QAE in anomaly detection.

\section{Summary \& Conclusion}\label{sec:conclusion}

In this study, we investigated the impact of different data embedding techniques on enhancing the representability of a quantum autoencoder for anomaly detection. We focused on three distinct embedding approaches: data reuploading, parallel embedding, and alternate embedding, comparing their effects against a standard ansatz composition. Our primary objective was to highlight the importance of effectively mapping data onto a higher-dimensional manifold instead of relying solely on complex and costly ansatz constructions.

We employed well-established, strongly correlated layers across all our benchmarks for the variational ansatz. Our findings demonstrate that, regardless of the data's dimensionality, the choice of data embedding technique significantly influences the quality and representability of the variational quantum algorithm for anomaly detection. Notably, even a modest increase in the local Hilbert space—from 2 to 4 dimensions for a single feature—resulted in substantial improvements. This suggests that further enlarging the data embedding space could lead to even more pronounced enhancements in performance.

However, it is essential to acknowledge that such data embedding methods we explored, mainly parallel and alternate embeddings, entail higher resource costs due to the increased number of qubits required. This resource intensiveness shows that these techniques may be most practical in the fault-tolerant era of quantum computing, where qubit availability and coherence times are no longer limiting factors.

In conclusion, our study highlights the critical role of data embedding in optimizing the performance of quantum autoencoders for anomaly detection. As quantum hardware continues to evolve, adopting more sophisticated embedding techniques will likely be vital to unlocking the full potential of quantum machine learning algorithms, enabling them to tackle increasingly complex and high-dimensional data with greater accuracy.

\section{Dataset and code availability}\label{sec:reproduce}

The code used for this analysis can be found in \href{https://github.com/jackaraz/qvae-anomaly}{this GitHub repository}. The credit card fraud data has been retrieved from \href{https://www.kaggle.com/datasets/mlg-ulb/creditcardfraud?resource=download}{Kaggle website}, and 2D datasets are generated using \textsc{Scipy} (version 1.10.1)~\cite{2020SciPy-NMeth}. 

\section{Acknowledgement}

JYA is supported by the U.S. Department of Energy, Office of Science, Office of Nuclear Physics with contract No.~DE-AC05-06OR23177, under which Jefferson Science Associates, LLC operates Jefferson Lab and in part by the DOE, Office of Science, Office of Nuclear Physics, Early Career Program under contract No. DE-SC0024358.

\bibliography{bibliography}
\end{document}